\documentclass[aps,prl,twocolumn,groupedaddress]{revtex4-2}

\usepackage{graphicx}
\begin{document}


\title{A new channel for excitonic transport in the FMO complex}


\author{Daniel L{\'o}pez D{\'i}az}\email{danielld@icimaf.cu}
\author{Gabriel Gil}\email{gabriel@icimaf.cu}
\author{Augusto Gonz{\'a}lez}\email{agonzale@icimaf.cu}
\affiliation{Institute of Cybernetics, Mathematics and Physics, Havana, Cuba}



\date{\today}

\begin{abstract}
We expose a new excitation transport channel in FMO, consisting of the 7 intramonomeric pigments and their 2nd nearest intermonomeric counterpart. Such a channel outcompetes the standard alternative, where intramonomeric pigments partner with their closest intermonomeric pigment. The efficient performance of the new channel lies at the interplay between incoherent energy transfer and the capacity of quantum coherence to build optimal acceptor levels. Finally, the differential fitness and peculiarity of the new over the alternative channel highlights its possible role within natural selection.
\end{abstract}


\maketitle

Quantum biology was pioneered in the 1940s by Schr\"odinger with the publication of `What is life?' \cite{schr1944}. Ever since, interests in the field have revolved around the mechanisms behind vision, olfaction, enzymatic activity and bird navigation \cite{lambert2013}. However, in the last decade, quantum biology has experienced a new boost due to the collection of evidence regarding the presumptive \cite{collini2013,cao2020} role of quantum coherence in the optimal efficiency of photosynthesis. Particularly, experimental results suggest the wave-like nature of the excitation energy transport that takes place within a key photosynthetic unit of green sulfur bacteria \cite{engel2007,panit2010}, namely, the Fenna-Matthews-Olson (FMO) protein-pigment complex \cite{fenna1975,olson2005}.

\begin{figure}[h!]
	\centering
	\includegraphics[width=0.6\linewidth]{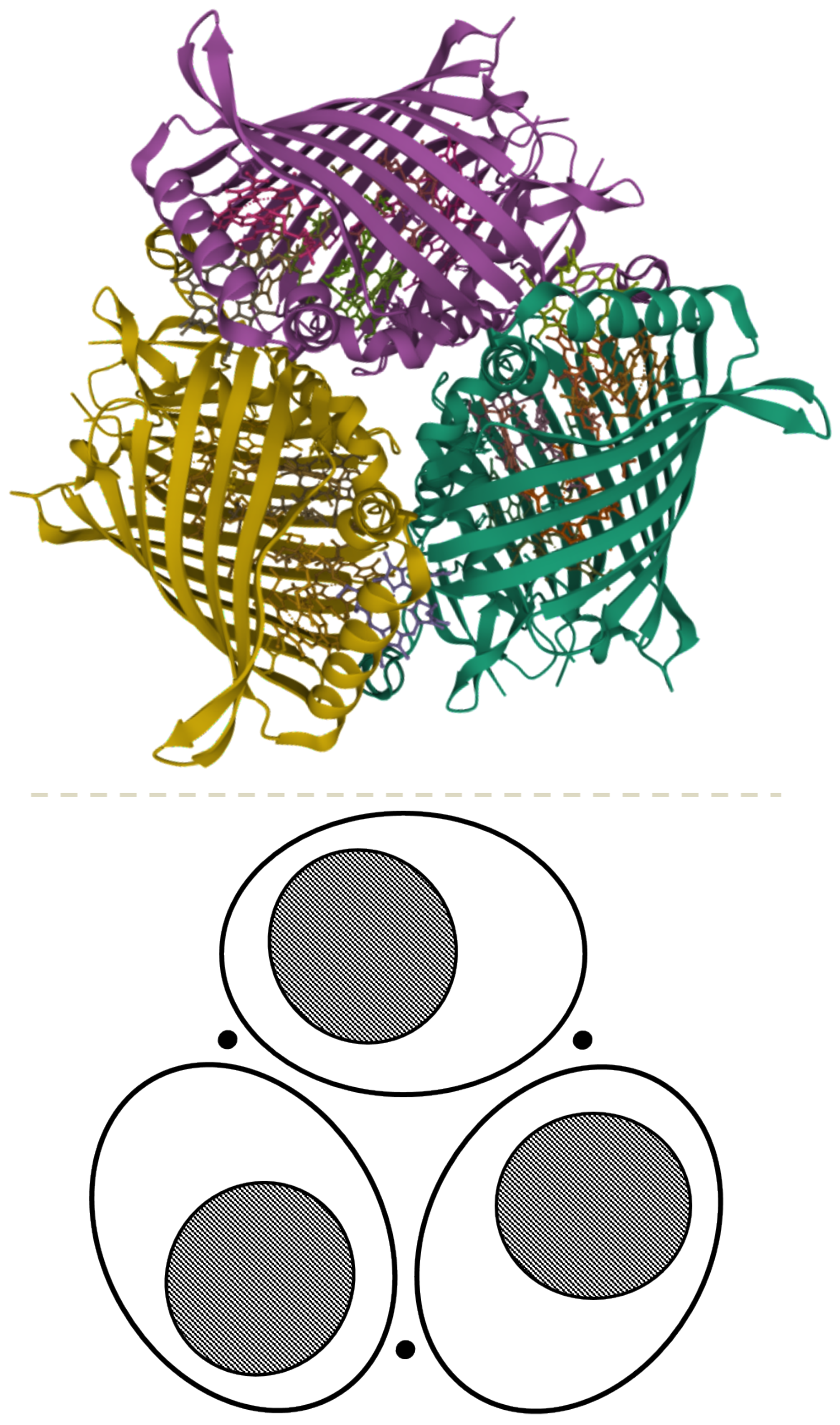}
	\caption{Top: 3D structure of the FMO trimer taken from Ref.~[\onlinecite{pdb3eni2009}]. Bottom: Associated schematics, highlighting the $C_3$ symmetry of FMO trimer (normal to the visual plane). Protein monomers are represented as empty solid-lined ellipses, whereas intra and intermonomeric pigments are shaded circles within them and dots outside them, respectively.}
	\label{fig:0}
\end{figure}

The FMO acts as a natural bridge allowing for the sunlight energy harvested at the chlorosomes, the antenna system of green sulfur bacteria, to be delivered to the reaction center, where a new photosynthesis stage starts. The FMO is composed of a protein trimer with $C_3$ symmetry (see Figure~\ref{fig:0}). Overall, such a trimer hosts 24 Bacteriochlorophyll a pigments (eight per monomer). Seven of such (for each monomer) are embedded in the protein scaffold, whereas the remaining three pigments are located at the surface of and in between monomeric units. Henceforth, we refer to the former and the latter as intra and intermonomeric pigments, respectively.

Until 2009, experimental drawbacks during the obtention of isolated monomers of FMO prevented the discovery of the intermonomeric pigments \cite{tronrud2009}. Due to their relative vicinity to the antenna system, it is currently assumed that these pigments receive the just-harvested solar energy and initiate the excitation transport among FMO pigments. Remarkably, the first report of long-time coherences in the excitation transport of FMO concerns a single monomer lacking intermonomeric pigments \cite{engel2007}.

The excitation transport in the trimer does not involve intramonomeric pigments from different FMO monomers \cite{milder2010}. Therefore, FMO monomers are functionally independent. However, the functional unit is still undetermined since it is unclear which group of intramonomeric pigments each intermonomeric pigment interacts with. In fact, the arrangement of the two intermonomeric pigments which are adjacent to a monomer is not symmetrical with respect to the intramonomeric pigments, making one (and not the other) significantly more closer and more strongly coupled to the intramonomeric pigments in question (see Figure~\ref{fig:0}). Consequently, the previous underdetermination is often resolved by appealing to geometric and energetic intuitions \cite{jia2015}.

We suggest that a \emph{counterintuitive} functional unit of FMO, where intramonomeric pigments (labeled 1-7) couple to their second nearest intermonomeric counterpart (labeled 8), can outcompete the more \emph{intuitive} one, where intramonomeric pigments associate with the closest intermonomeric pigment (labeled as 8$^\prime$). The latter amounts to say that there is an efficient channel for the excitation out of pigment 8 passing through the second nearest (and not the nearest) group of intramonomeric pigments. In order to support this claim, we perform excitonic transport simulations for both (independent) functional units of FMO, hereafter FMO8 and FMO8$^\prime$. 

FMO8$^\prime$ is described by a fully coherent, quantum excitation transport \cite{zech2014}. Choosing an incoherent over a coherent transport mechanism for FMO8$^\prime$ delivers qualitatively similar results as compared with the transport through the FMO8 channel (see Supplemental Material for a detailed discussion). As for FMO8, we rely on a \emph{hybrid} mechanism that combines quantum coherence, known to be active for the intermonomeric pigments \cite{panit2010}, with a more realistic F\"orster Resonance Energy Transfer (FRET) \cite{forster1948} out of pigment 8. FRET is often considered between two pigments, acting as a donor and acceptor of the excitation. In the FRET mechanism we are proposing, pigment 8 donates energy to a collective excitonic state of pigments 1-7.

The situation for the FMO8 is akin to that of the LH2 protein-pigment complex of purple bacteria, where two excitonic states delocalize over disjoint bacteriochlorophyll rings and act as donor and acceptor levels of a \emph{generalized} FRET process \cite{caprasecca2018}. However, the FMO8 depend only upon a delocalized acceptor (not donor) state. 

The excitonic Hamiltonian is taken from Ref.~[\onlinecite{jia2015}]. For the hybrid mechanism, we considered a (Gaussian) homogeneous broadening of 80 cm$^{-1}$ \cite{milder2010} and a Stokes shift of 18 cm$^{-1}$ \cite{ratsep2007}, in order to compute the spectral overlap between pigment 8 and any excitonic state involving pigments 1-7. 

To assess the excitonic transport efficiency of FMO8 and FMO8$^\prime$, we choose the time-average probability \cite{mohseni2008} of finding the excitation at pigment 3, which is the closest to the reaction center \cite{milder2010}. Our exploration is limited to short times (up to 125 fs), which goes beyond recent estimations of the quantum coherence timespan in FMO (i.e., $\sim$60 fs) \cite{duan2017}. 

\begin{figure}[h!]
	\centering
	\includegraphics[width=\linewidth]{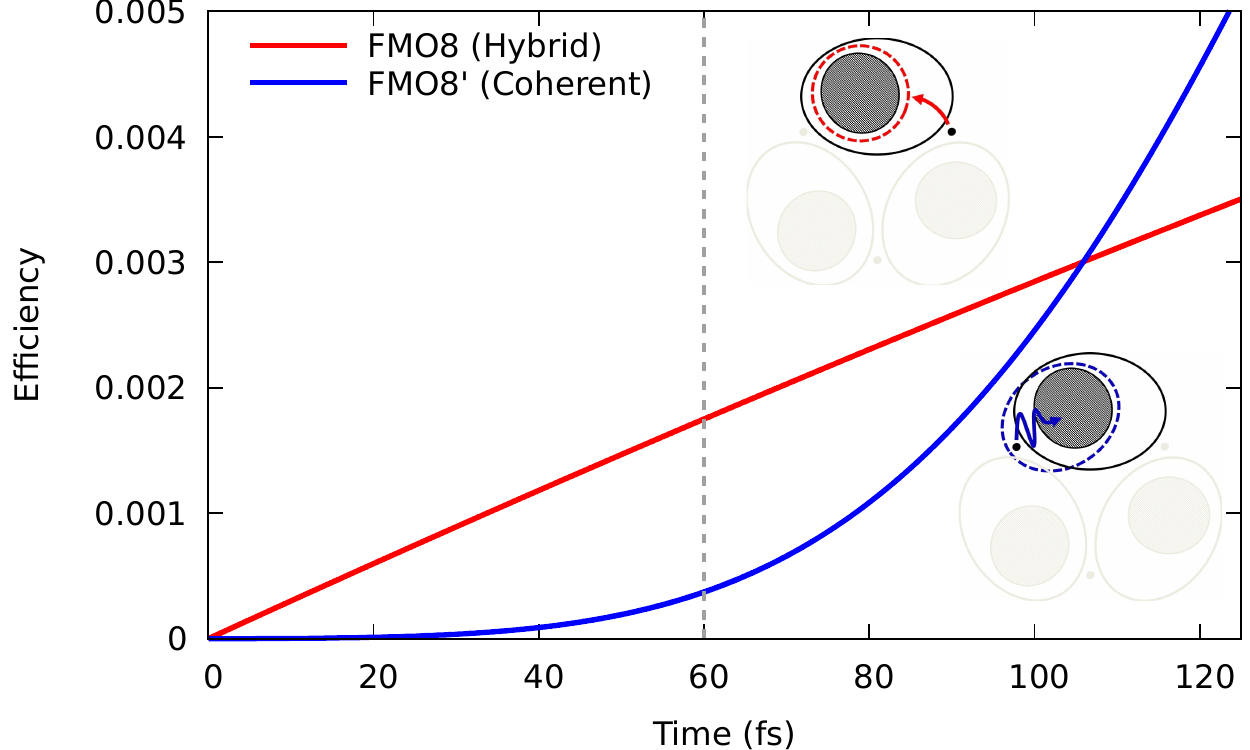}
	\caption{Efficiency vs. time for FMO8 and FMO8$^\prime$. The FMO8$^\prime$ uses a quantum coherent evolution (blue) whereas the FMO8 uses the hybrid mechanism (red). Inset: each curve is partnered with a cartoon depicting the full FMO trimer, while emphasizing (in black and white) the involved subsystem (whether FMO8 or FMO8$^\prime$) and the excitonic transport mechanism at work (whether hybrid or coherent). FRET (quantum evolution) is signified with an arched (zigzag) arrow, and the quantum coupling region lies within dashed elliptical contours. Other schematical elements are shared with Figure~\ref{fig:0}.}
	\label{fig:1}
\end{figure}

Results are shown in Figure \ref{fig:1}. Remarkably, we find that FMO8 efficiency is greater than that of FMO8$^\prime$ for the first $\sim$100 fs, particularly five times so for 60 fs. Furthermore, note that the overturn takes place far from the quantum coherence time window. These findings underpin the previously overlooked FMO8 as the operative excitation transport channel within the FMO.

The reason for the better performance of FMO8 is manifold. On the one hand, (hetero) FRET is an inherently asymmetric process upon the swapping of energy donor and acceptor. This is ultimately due to the Stokes shift. Indeed, FRET is favored when the absorption band of the acceptor is less energetic than the absorption band of the donor, and it is disfavored otherwise \cite{forster1948}. As a consequence, there is an advantageous restriction to the excitation back-transfer built in the hybrid mechanism of FMO8, where pigments 1-7 act (collectively) as an acceptor and pigment 8 as a donor. Conversely, the quantum coherent transport is equally optimal in both directions, from input to output and vice versa. In fact, for a pair of two-level systems, the coherent transport efficiency only depends on the absolute difference of excitation energies \cite{cohen1991}. On the other hand, the hybrid mechanism in FMO8 seems to capitalize on the coherent coupling of the intramonomeric pigments in different ways: (a) by tuning the acceptor level slightly below resonance with pigment 8, in order to improve the spectral overlap associated with forward (but not backward) FRET; (b) by shifting the ``center-of-mass" of the acceptor state in the direction of pigment 8; and finally, (c) by enhancing and (d) aligning the acceptor transition dipole with respect to pigment 8.

The effect of (a), (b), (c) and (d) is apparent in Table \ref{tab:1}. Therein, for each excitonic state $i$ of the pigments 1-7, we list: the spectral overlap of a donor pigment 8 (boldfaced subindex) and an acceptor state $i$ minus the converse ($\Delta J_{\mathbf{8}i}$), the distance from pigment 8 to the center-of-mass of $i$ ($r_{\mathbf{8}i}$), the transition dipole magnitude of $i$ ($d_i$), and the orientational factor with reference to $i$ and pigment 8 ($\kappa_{\mathbf{8}i}^2$). 

We compute the dipole as $d_i=\left|\sum_\alpha C_\alpha^{(i)}\mathbf{d}_\alpha\right|$ and the distance as $r_{\mathbf{8}i}=\left|\mathbf{r}_8-\sum_\alpha|C_\alpha^{(i)}|^2\mathbf{r}_\alpha\right|$, where $\mathbf{r}_\alpha$ and $\mathbf{d}_{\alpha}$ are the position and transition dipole of pigment $\alpha$, and $C_\alpha^{(i)}$ is the excitonic coefficient of pigment $\alpha$ (1-7) in the state $i$. In the dipolar approximation, the rate difference of a FRET outgoing from pigment 8 and another incoming to it, reads $\Delta k_{\mathbf{8}i}=2\pi\kappa_{\mathbf{8}i}^2~d_i^2~d_\mathbf{8}^2~\Delta J_{\mathbf{8}i}/r_{\mathbf{8}i}^6$ \cite{forster1948}. 

Pigment's positions and dipole orientations are obtained as in Ref.~[\onlinecite{gil2018}], provided the FMO structure from \textit{Chlorobaculum Tepidum} \cite{pdb3eni2009}. We fit an effective (pigment) dipole magnitude ($\sim$4.6 D) in order to coordinate the off-diagonal Hamiltonian matrix elements of Ref.~[\onlinecite{jia2015}] with the corresponding dipolar couplings.

Clearly, the 4th excitonic state is the optimal acceptor for a FRET involving pigment 8 as a donor. First, it exhibits the highest $\Delta J_{\mathbf{8}i}$ as pointed out in (a). Secondly, such a state features the second smallest $r_{\mathbf{8}i}$, indicating a tendency of the delocalized excitation to get as close as possible to pigment 8, in connection with (b). Moreover, the 4th excitonic state has the greatest transition dipole ($\sim 8.2$ D), and so it amplifies the single-pigment effective dipole in the sense of (c). In fact, such a state is a dipole resonance \cite{capote2001}, taking up as much as 42$\%$ of the energy-weighted sum $\sum_i E_i d_i^2$. In addition, the orientational factor $\kappa_{\mathbf{8}i}^2$ corresponding to the 4th excitonic state, albeit non-optimal, is second largest among the seven excitonic states, supporting (d). Notably, $r_{\mathbf{8}i}$ and $\kappa^2_{\mathbf{8}i}$ values for the 4th excitonic state are indeed optimal within the $\Delta J_{\mathbf{8}i}>0 $ constraint --meaning that the 4th stand out from the possible acceptor states of pigments 1-7 in all relevant respects. 

\begin{table}[h!]
	\centering
	\begin{tabular}{cccccccc}
		\hline
		\hline
		$i$ & $\Delta J_{\mathbf{8}i}$ (au) & $r_{\mathbf{8}i}$ ($\AA$) & $d_i$ (D) & $\kappa^2_{\mathbf{8}i}$\\
		\hline
		1 & 2.13 & 34.41 & 4.387 & 0.550\\
		2 & 88.46 & 28.75 & 5.232 & 0.158\\
		3 & 295.68 & 36.11 & 5.498 & 0.193\\
		$~$4* & 705.42 & 26.78 & 8.287 & 0.715\\
		5 & -547.17 & 25.70 & 1.721 & 1.032\\
		6 & -251.08 & 33.06 & 3.462 & 0.021\\
		7 & -164.27 & 26.91 & 1.911 & 0.411\\
		\hline
		\hline
	\end{tabular}
	\caption{Ingredients entering FRET rate from pigment 8 to an excitonic state of pigments 1-7. From left to right, we show the state index $i$, the spectral overlap difference $\Delta J_{\mathbf{8}i}$, the distance $r_{\mathbf{8}i}$, the transition dipole $d_i$ and the orientational factor $\kappa_{\mathbf{8}i}^2$. A star highlights the optimal state.}
	\label{tab:1}
\end{table}

\begin{figure}[h!]
	\centering
	\includegraphics[width=0.65\linewidth]{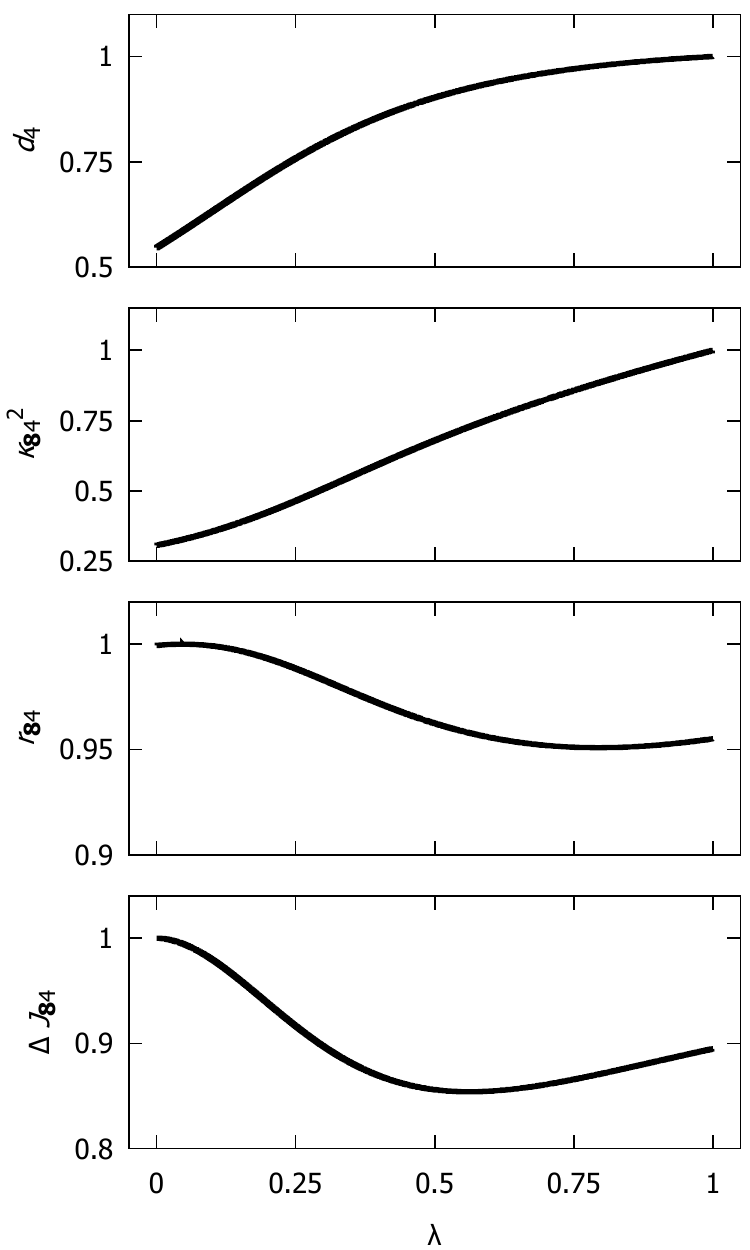}
	\caption{FRET ingredients as a function of a coupling parameter $\lambda$ (normalized to the highest value in range). From top to bottom, we show the behavior of the spectral overlap difference $\Delta J_{\mathbf{8}4}$, the distance $r_{\mathbf{8}4}$, the transition dipole $d_4$ and the orientational factor $\kappa_{\mathbf{8}4}^2$.}
	\label{fig:2}
\end{figure}

The coupling within intramonomeric pigments is in fact responsible for the optimization of the FRET from pigment 8 to the 4th excitonic state of pigments 1-7. To support such a claim, we introduce an \emph{artificial} coupling parameter $\lambda$ modulating the interaction among intramonomeric pigments from the non-interacting ($\lambda=0$) to the fully-interacting ($\lambda=1$) regime. Importantly, the interaction with pigment 8 is $\lambda$-independent. Figure~\ref{fig:2} shows every FRET rate ingredient as a function of $\lambda$. As it is apparent, the dipole and orientational factor increase monotonously with the interaction strength. On a different note, there is not much room for an improvement of the distance, which is strongly constrained by the roughly planar surface where the pigments 1-7 rest. However, we observe a small decrease of $r_{\mathbf{8}4}$ below the minimal inter-pigment distance from $\alpha$ to 8. The case of the spectral overlap difference is special because the 3rd and 4th excitonic levels are exactly degenerate at $\lambda=0$ (since pigment-4's and 5's energy coincide \cite{jia2015}). Such degeneracy causes the 4th excitonic level to ramp up quickly toward the resonance with pigment 8 upon increasing $\lambda$ (not shown). This energy increase is later on counteracted by a stabilization under the influence of other pigments, enhancing $\Delta J_{\mathbf{8}4}$ near the fully-interacting regime.

The fact that the 4th excitonic state stands out from the rest in all relevant respects is highly unexpected since all ingredients entering FRET rate are independent from each other. This peculiarity follows from the special configuration of the pigments within the FMO, which is ultimately tailored by natural selection.

At this point, the question arises as to which of the two competing FMO excitation channels (namely, FMO8 and FMO8$^\prime$) could have been naturally selected. Indeed, we expect that, during the course of biological evolution of green sulfur bacteria, the position and orientation of FMO pigments underwent a long optimization toward high excitonic transport efficiencies, therefore contributing to the better and better fitness the host organisms. Now, if the FMO configuration have been preferentially selected to enhance FMO8 (and not FMO8$^\prime$) performance, we can add up important evidence in favour of FMO8 as the biologically operating channel. 

To develop some insight from natural selection, we rely on a genetic algorithm acting on a set of initially random configurations of eight pigments. Pigments are assumed to be point-like transition dipoles as before. Hence, configurations are solely described by their excitation energies, positions and orientations. \emph{A priori}, we are looking for optimized configurations that are not so different to FMO8 and FMO8$^\prime$. To that aim, we perform independent simulations for the two cases, where the \emph{ancestral} generation is uniformly sampled within a collection of spheres centered at either FMO8 or FMO8$^\prime$ actual positions. Sampling sphere radii are taken as half the distance to the closer sphere center, in order to grant as much variability as possible, while keeping the coarse-grained geometry of either FMO8 or FMO8$^\prime$. We reject configurations with pigments that come closer than one-tenth the minimum distance between actual FMO pigments. Dipole orientations are uniformly sampled without constraints. Pigment excitation energies $E_{\alpha}$ are sampled from a Gaussian distribution with the mean and standard deviation of actual FMO energies \cite{jia2015}. Random values beyond the three-sigma limit are always rejected. It is worth noting that in the ancestral generation, each FMO8-like configuration has an homologous FMO8$^\prime$-like configuration sharing excitation energies, positions and dipole orientations of intramonomeric pigments, and differing only in what concerns to pigment 8 and 8$^\prime$.

Every \emph{further} generation is obtained from a mutation and a selection process. Mutations apply a Gaussian noise to configurations in the current generation (parents). The noise preserves mean values from energies, positions and orientations in the current generation. Standard deviations are however down scaled by the number of past generations, so as to mimic the effect of accumulated adaptations. The initial value of such standard deviations is one-tenth that of the ancestors. Ten mutated (child) configurations are generated per parent (reproduction), and only a child per parent survives to be reproduced. Survivor selection is grounded on its excitonic transport efficiency at 60 fs (cf.~Fig.~\ref{fig:1}). Efficiency is computed as in either the hybrid or the coherent mechanism, depending on whether our coarse-grained geometry is FMO8-like or FMO8$^\prime$-like. Such a genetic algorithm takes inspiration from Ref.~[\onlinecite{walschaers2013}], while incorporating further degrees of freedom (pigments' excitation energies and positions) to the evolution of the FMO-like configurations.

\begin{figure}[h!]
	\centering
	\includegraphics[width=0.75\linewidth]{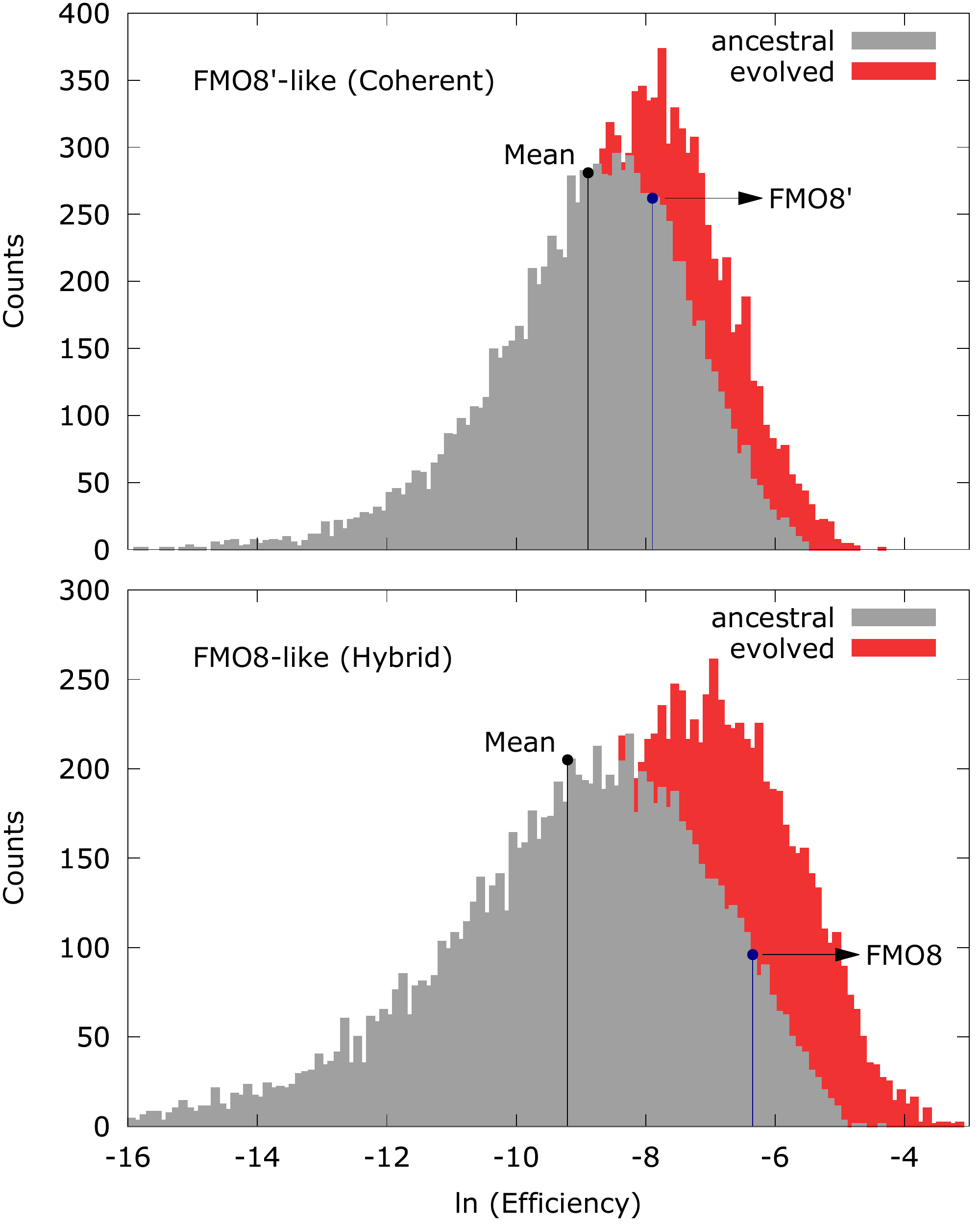}
	\caption{Histograms of the logarithm (base $e$) of the efficiency for $10^4$ FMO8$^\prime$-like (top) and FMO8-like (bottom) random configurations. The ancestral and the evolved configurations are displayed in gray and red, respectively. Configurations are better fitted the more efficient it is the underlying mechanism for excitonic transport (which is coherent for FMO8$^\prime$-like, and hybrid for FMO8-like systems). In the top (bottom) panel, the actual FMO8$^\prime$ (FMO8) efficiency  and the reference efficiency of ancestral FMO8$^\prime$-like (FMO8-like) systems are indicated by vertical lines (topped with a dot) in blue and black, respectively. The position of the dot in the vertical axis correspond to the nearest bin count within the ancestral generation.} 
	\label{fig:3}
\end{figure}

Figure \ref{fig:3} shows the resulting efficiency histograms for  FMO8$^\prime$-like and FMO8-like configurations in the ancestral and an \emph{evolved} stage after 100 generations. Notice that the scale-free behavior of efficiency distributions suggests the log scale representation of eficiency histograms, and the geometric mean as a reference value for the efficiency of a given generation. As expected, the evolved is skewed toward the right of the ancestral generation for both, top and bottom panels. 

On a basis of a logistic fit of the reference efficiency as a function of the first 100 generations (not shown), we emphasize: (i) that the reference efficiency of FMO8$^\prime$-likesystems reaches that of actual FMO8$^\prime$ within 622 generations circa, and (ii) that actual FMO8 still outperforms the \emph{most} evolved FMO8-like systems with (aymptotic) reference efficiency. As a consequence, natural selection of FMO seems to be more inline with the operation of FMO8 channel than with the currently accepted alternative (namely, FMO8$^\prime$). Indeed, FMO is better fitted and more peculiar within the phase space of FMO8 evolution. We estimate the fitness $f$ as the relative difference between actual FMO8 (FMO8$^\prime$) efficiency and the reference efficiency of ancestral FMO8-like (FMO8$^\prime$-like) systems. On the other hand, the peculiarity $p$ of FMO8 (FMO8$^\prime$) within ancestral FMO8-like (FMO8$^\prime$-like) configurations is quantitatively captured by the relative difference of bin counts at the aforementioned reference efficiency and at the actual FMO8 (FMO8$^\prime$) efficiency.  Ingredients entering $f$ and $p$ estimations are highlighted by vertical lines in Fig.~\ref{fig:3} (see caption). Indeed, we get $f=94\%$ and $p=53\%$ for FMO8, while $f=63\%$ and $p=7\%$ for FMO8$^\prime$, confirming our qualitative description. Note that high $f$ and $p$ is what is expected from a biological function that have evolved through natural selection. That is, due to a reiterated selection process, today's FMO should be \emph{extremely} better fitted than typical FMO-like systems in the ancestral generation. Moreover, today's FMO could only have arisen out of pure chance in the ancestral generation due to an \emph{extremely} rare (peculiar) mutation.

Our results suggests at least a competition between two different channels for the excitation out of the superficial pigments and towards the reaction center. To this end, we have performed simulations focusing on each independent functional units of FMO (i.e., FMO8 and FMO8$^\prime$) or alike. However, in order to pinpoint \emph{definitively} the actual excitation channel operating in FMO, we need (I) to expand the simulation cell to cover at once the FMO8 and FMO8$^\prime$ and (II) to rely on a unified framework, such as open quantum system theory \cite{zech2014}, including the coherent and incoherent mechanisms (i.e., master eq. vs. Schr\"odinger eq.) as limiting cases. A Lindblad model for the FMO trimer will be presented elsewhere to tackle these important issues.

In summary, we have exposed a new counterintuitive channel for the excitation transport within FMO that has been previously overlooked, namely, the FMO8. The FMO8 is composed of the seven intramonomeric pigments of FMO and a non-trivial intermonomeric pigment which is the \emph{second} closest to the latter. Our simulations show that FMO8 is superior to the standardly explored channel for the excitation transport in FMO, referred to as FMO8$^\prime$. We argue that the efficient operation of FMO8 relies on the interplay between an incoherent energy transfer from the intermonomeric pigment 8 and a quantum coherent state encompassing the intramonomeric pigments 1-7. Finally, a biologically inspired view allow us to further substantiate our claims about the relevance of FMO8.

\begin{acknowledgements}
\textbf{Acknowledgements} This work was supported by the project No. NA211LH500-002 from AENTA-CITMA (Cuba), and the Caribbean Network for Quantum Mechanics, Particles and Fields (ICTP). Preliminary results were presented as a short talk at the FT45 workshop, in the ocassion of the 45th Anniversary of the Theoretical Physics Department of ICIMAF (Havana, Cuba), and as a seminar to the Nanostructures and (Bio)molecules Modeling Group from the University of Padua (Italy). Authors are especially grateful to R. Mulet, A. Cabo, H. P{\'e}rez, S. Corni, A. Migliore and F. Grasselli for useful discussions.
\end{acknowledgements}

\bibliography{Paper}

\end{document}



\title{Supplemental Material: A new channel for excitonic transport in FMO}


\author{Daniel L{\'o}pez D{\'i}az}\email{danielld@icimaf.cu}
\author{Gabriel Gil}\email{gabriel@icimaf.cu}
\author{Augusto Gonz{\'a}lez}\email{agonzale@icimaf.cu}
\affiliation{Institute of Cybernetics, Mathematics and Physics, Havana, Cuba}



\date{\today}

\begin{abstract}
\end{abstract}


\maketitle

\section{A. Other possible mechanisms for FMO8 and FMO8$^\prime$}
In the main text, we tackled the FMO8 and FMO8$^\prime$ excitation transport via a hybrid (incoherent-coherent) and a fully coherent mechanism, respectively. Albeit, there are other possible excitation transport for FMO8 and FMO8$^\prime$. For example, FMO8 or FMO8$^\prime$ could make use of a fully incoherent (FRET-like) excitation transfer between pigments, despite the evidence \cite{duan2017}. On the other hand, however unlikely due to the small coupling between the input site and the rest of the system, the excitation transport within FMO8 may also undergo a fully coherent (wave-like) evolution from pigment 8 to pigments 1-7.

\renewcommand{\thefigure}{S1}
\begin{figure}[h!]
	\centering
	\includegraphics[width=0.65\linewidth]{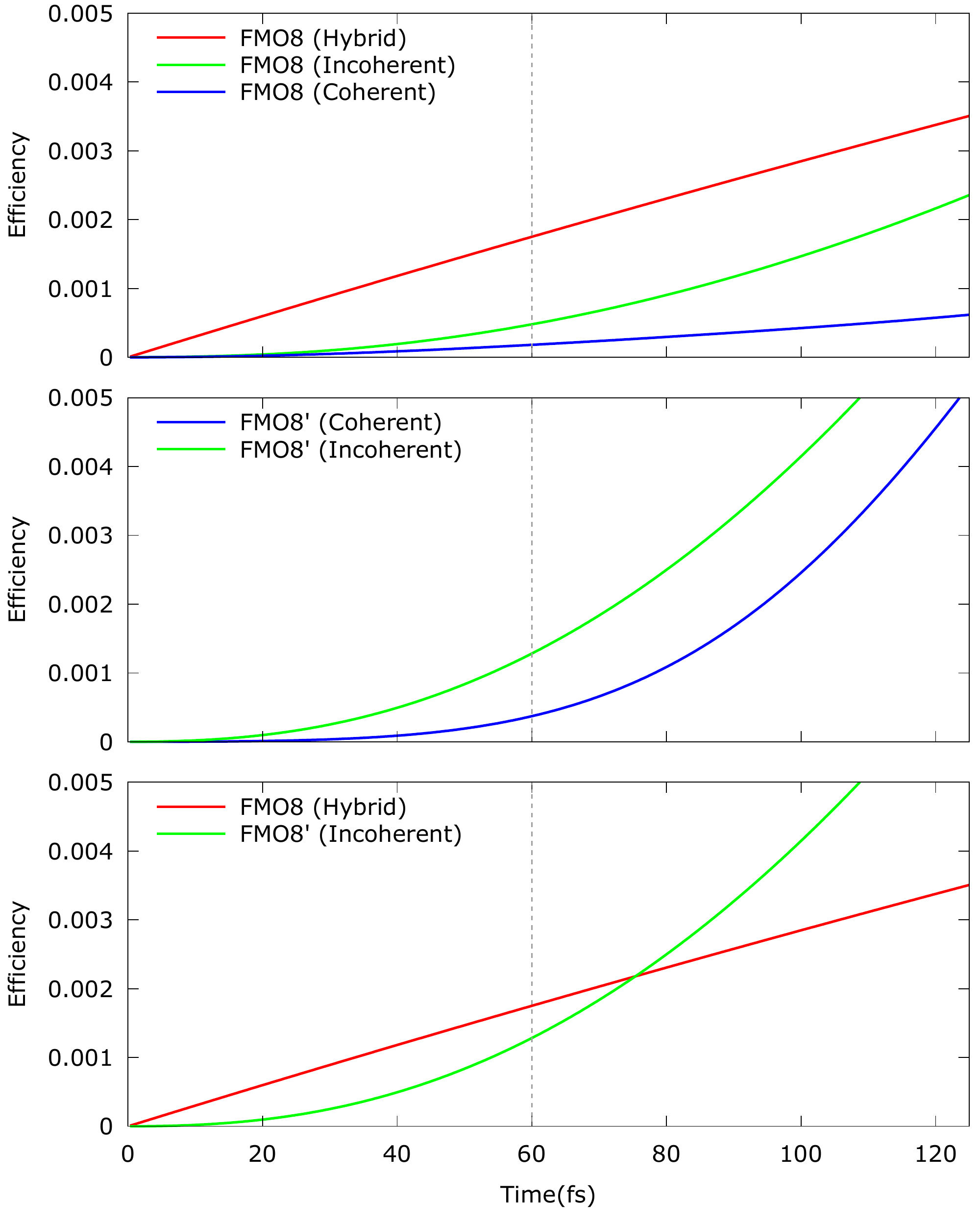}
	\caption{Efficiency vs. time for FMO8 and FMO8$^\prime$ systems. In the top (middle) panel we show the 3(2) possible mechanisms for FMO8 (FMO8$^\prime$) channel, whereas in bottom panel we combine the best mechanism for FMO8 and FMO8$^\prime$ channels. In all panels, a green (blue) curve corresponds to an incoherent (coherent) mechanism, whereas a red curve correspond to a hybrid mechanism.}
	\label{fig:s1}
\end{figure}

In the first panel of Fig.~\ref{fig:s1}, we show the efficiency of a fully coherent, a fully incoherent and a hybrid mechanism for FMO8. In the timeframe explored (125 fs), the best mechanism for FMO8 is the hybrid, followed by a fully incoherent transport. The second panel of Fig.~\ref{fig:s1} shows analogous results for FMO8$^\prime$. As apparent, the incoherent mechanism outperforms the coherent one, being roughly three times as much efficient at 60 fs. 

Finally, in the third panel of Fig.~\ref{fig:s1}, we contrast the best mechanisms both channels, namely, the hybrid mechanism for FMO8 and the fully incoherent mechanism of FMO8$^\prime$. As compared with Fig. 1 (main text), Fig.~\ref{fig:s1} is qualitatively similar. Although the overturn is closer in time (~80 fs), FMO8 is favored over FMO8$^\prime$ in the short term. Additionally, the gap between the hybrid FMO8 and the incoherent FMO8$^\prime$ efficiency curves is narrower (cf. Fig. 1, main text), making the FMO8$^\prime$ incoherent being three-quarters as efficient as the FMO8 hybrid at 60 fs.

It is noteworthy that a hybrid mechanism for FMO8$^\prime$ is inconsistent, and therefore physically irrelevant. The hybrid mechanism we are proposing requires the weak coupling between the intermonomeric pigment in question and the rest of the system. This condition is met by FMO8 but not FMO8$^\prime$. In fact, pigment 8 is approximately twice as far from the closest pigment as pigments 1-7 between each other. Conversely, the distance between pigment 8$^\prime$ and its nearest pigment is roughly the same as that between intramonomeric pigments. However, such intuitions may be unreliable since we are not yet considering the orientation factor of the dipole-dipole coupling between pigments. 

To ascertain that FMO8$^\prime$ does not meet the requirement of the hybrid mechanism, we analyze quantitatively the Hamiltonian, which synthesize both distance and orientation factors. The Hamiltonian of \cite{jia2015} is tabulated in Tab.~\ref{tab:s1} to facilitate the reading. Notice that the interaction between pigment 8$^\prime$ and certain intramonomeric pigments (e.g., pigment 1) is of the same order as the interaction between intramonomeric pigments. Therefore, we cannot treat inter-intramonomeric transfer as incoherent while considering intramonomeric pigments as coherently coupled. 

\renewcommand{\thetable}{SI}
\begin{table}[h!]
	\centering
	\begin{tabular}{|c|c|c|c|c|c|c|c|c|c|}
		\hline
		\hline
		Pigments & 1 & 2 & 3 & 4 & 5 & 6 & 7 & 8 & 8$^\prime$\\
		\hline
		1 & 11550 & -91.0 & 4.1 & -6.3 & 6.3 & -8.8 & -7.8 & 0.2 & 32.4\\
		\hline
		2 & $-$ & 11413 & 28.7 & 8.2 & 1.0 & 8.8 & 3.4 & 0.9 & 6.3\\
		\hline
		3 & $-$ & $-$ & 11332 & -46.6 & -4.4 & -9.3 & 1.3 & 2.2 & 1.3\\
		\hline
		4 & $-$ & $-$ & $-$ & 11437 & -73.9 & -17.7 & -59.1 & -2.0 & -1.9\\
		\hline
		5 & $-$ & $-$ & $-$ & $-$ & 11437 & 76.0 & -3.1 & 6.1 & 4.2\\
		\hline
		6 & $-$ & $-$ & $-$ & $-$ & $-$ & 11518 & 25.9 & -4.2 & -11.6\\
		\hline
		7 & $-$ & $-$ & $-$ & $-$ & $-$ & $-$ & 11501 & -7.8 & -11.9\\
		\hline
		8 & $-$ & $-$ & $-$ & $-$ & $-$ & $-$ & $-$ & 11486 & $-$\\
		\hline
		~8$^\prime$ & $-$ & $-$ & $-$ & $-$ & $-$ & $-$ & $-$ & $-$ & $-$\\
		
		\hline
		\hline
	\end{tabular}
	\caption{Site energies (diagonal) and electronic couplings (off–diagonal) of FMO8 and FMO8$^\prime$ pigments in units of cm$^{-1}$, as taken from Ref.~[\onlinecite{jia2015}].}
	\label{tab:s1}
\end{table}

\section{B. Simulating natural selection of FMO8$^\prime$-like systems using incoherent excitation transport}

We are interested in the comparison between FMO excitation transport channels, not mechanisms. Hence, to complete our case for the FMO8 channel against FMO8$^\prime$, we need to supply a natural-selection argument for the inadequacy FMO8$^\prime$, even if the incoherent (instead of the coherent) excitation transport is active. To that aim, we run the genetic algorithm (see main text) for the FMO8$^\prime$ for an incoherent mechanism. 

\renewcommand{\thefigure}{S2}
\begin{figure}[h!]
	\centering
	\includegraphics[width=0.65\linewidth]{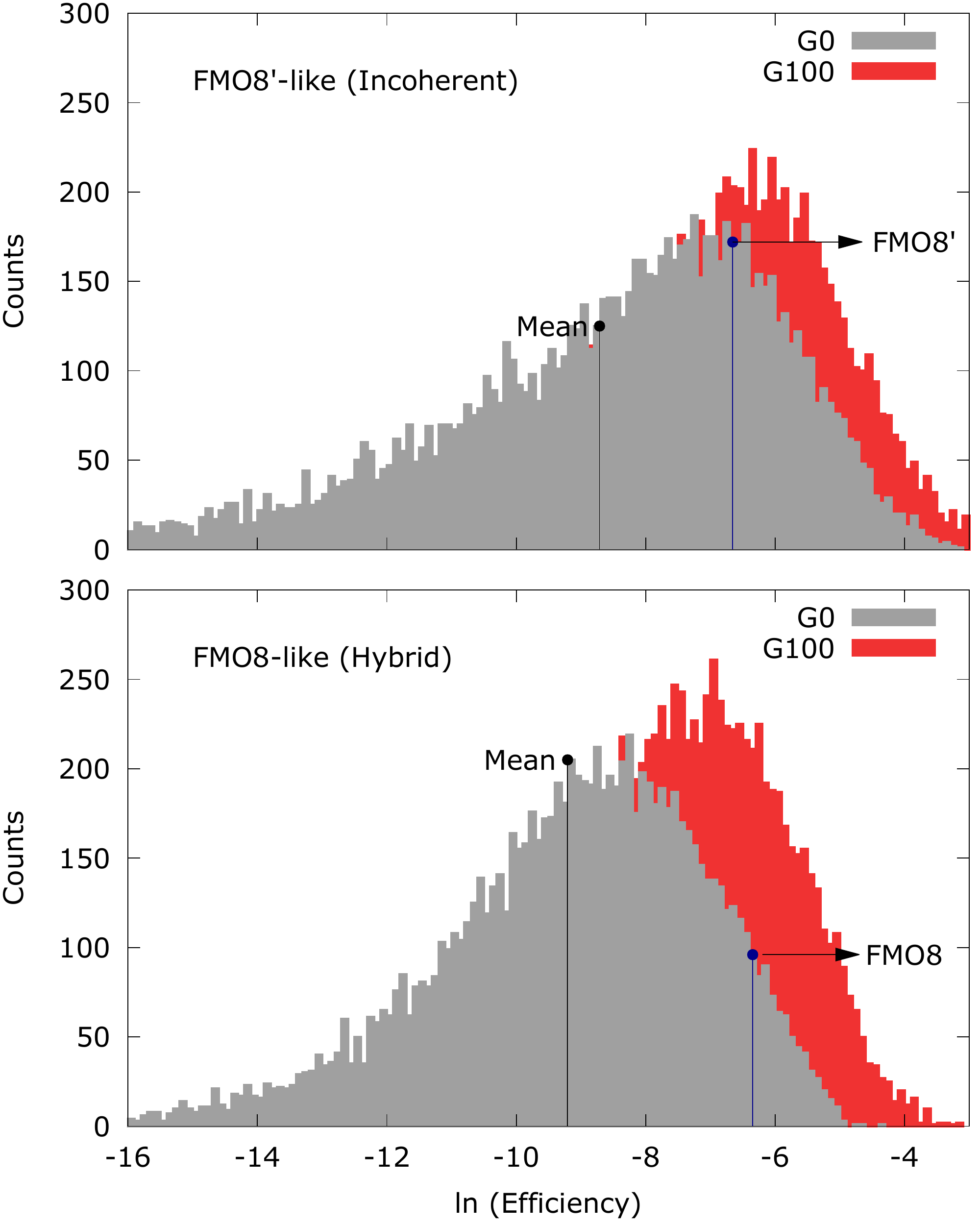}
	\caption{Histograms of the logarithm (base $e$) of the efficiency for $10^4$ FMO8$^\prime$-like (top) and FMO8-like (bottom) random configurations. The ancestral and the evolved configurations are displayed in gray and red, respectively. Configurations are better fitted the more efficient it is the underlying mechanism for excitonic transport (which is incoherent for FMO8$^\prime$-like, and hybrid for FMO8-like systems). In the top (bottom) panel, the actual FMO8$^\prime$ (FMO8) efficiency  and the reference efficiency of ancestral FMO8$^\prime$-like (FMO8-like) systems are indicated by vertical lines (topped with a dot) in blue and black, respectively. The position of the dot in the vertical axis correspond to the nearest bin count within the ancestral generation.} 
	\label{fig:s2}
\end{figure}

Results are shown in Fig.~\ref{fig:s2}. We replot the bottom panel of Fig. 3 (main text) to facilitate the comparison between FMO8 hybrid and FMO8$^\prime$ incoherent cases. Fig.~\ref{fig:s2} top shows the efficiency histograms for FMO8$^\prime$ incoherent excitation transport for the ancestral and an evolved generation after 100 steps. The behavior is akin to that for the coherent mechanism (see Fig. 3 top from main text). In particular, we note that the fitness and peculiarity of FMO8$^\prime$ within FMO8$^\prime$-like systems is significantly lower ($f=24\%$, $p=-42\%$) than those for FMO8 through the hybrid mechanism (see main text). Therefore, FMO8 also seems to be the biologically operating channel against the alternative FMO8$^\prime$, even if we consider a different transport mechanism for the latter.

\bibliography{Paper}


